\pdfoutput=1

\documentclass[11pt]{article}

\usepackage{acl}

\usepackage{times}
\usepackage{latexsym}
\usepackage{amsmath}
\usepackage{amssymb}
\usepackage[utf8]{inputenc}
\usepackage{graphicx}
\usepackage{lipsum}
\usepackage[justification=centering]{caption}
\usepackage{tabularx, multirow, booktabs}
\usepackage{wrapfig}

\usepackage[T1]{fontenc}

\usepackage[utf8]{inputenc}

\usepackage{microtype}

\title{DEFTri: A Few-Shot Label Fused Contextual Representation Learning For Product Defect Triage in e-Commerce}

\author{Ipsita Mohanty \\
  Walmart Global Tech \\
  Sunnyvale, California, USA\\
  \texttt{ipsita.mohanty@walmart.com} \\}

\begin{document}
\maketitle
\begin{abstract}
Defect Triage is a time-sensitive and critical process in a large-scale agile software development lifecycle for e-commerce. Inefficiencies arising from human and process dependencies in this domain have motivated research in automated approaches using machine learning to accurately assign defects to qualified teams. This work proposes a novel framework for automated defect triage (DEFTri) using fine-tuned state-of-the-art pre-trained BERT on labels fused text embeddings to improve contextual representations from human-generated product defects. For our multi-label text classification defect triage task, we also introduce a Walmart proprietary dataset of product defects using weak supervision and adversarial learning, in a few-shot setting.
\end{abstract}

\section{Introduction}
In large e-commerce organizations, there are many defects generated periodically with a massive pool of software teams and developers spread across geographies to pick from, each with unique domain specialization. Most organizations have a large pool of human triaging agents responsible for routing these product defects across various teams within the organization. However, large-scale software releases are time-sensitive, and effective defect assignments are a critical component in the process that is prone to bottlenecks. Determining the most suitable team to own a defect may require several attempts; thus, wasting time to diagnose a defect not in the team's domain of specialty and, overall, negatively impacting the defect resolution throughput. 

Prior industry research work on automated defect triage has primarily focused on using the traditional machine learning approaches. However, with the recent surge of state-of-the-art pre-trained language models, one under-explored field of application is operations in agile software development. In the defect triage, handling scenarios require Natural Language Understanding to utilize the context of the defects logged by human testers, to predict all the teams associated with resolution. The current defect triage process is primarily human-agents driven. This work integrates an automated defect triage framework, DEFTri using product defect's contextual features to achieve operational excellence within Walmart's software development lifecycle.

We propose a novel framework, DEFTri to perform an automated defect triage using contextual representations of human-generated defect texts. We use Walmart's proprietary data of product defects curated by product managers, program managers, and beta-testers to train our models. We use domain-specific lexicons to generate labeled training data using weak supervision in a few shot settings. We further use adversarial learning to increase our training sample size while increasing the robustness of our models. We propose our model architecture for fine-tuning pre-trained BERT \citep{DBLP:journals/corr/abs-1810-04805} for our multi-label classification task. Finally, we consolidate our experiments, analyze the results and discuss future research work.

\section{Related Work}
Prior research work on defect triage \citep{DBLP:journals/corr/abs-1910-05835, DBLP:journals/corr/abs-1801-01275,article} mostly focuses on using traditional machine learning and RNNs on word vector representations of text using BOW, Word2Vec, Tfidf, etc. Another recent research relies on graph representation learning for defect triage \citep{DBLP:journals/corr/abs-2101-11846}. This paper proposes a graph recurrent convolution network with a joint random walk mechanism-based architecture. Also, several recent research on label embedding \citep{XiongFWKO21, LIU2021385,DBLP:journals/corr/abs-2006-11991} has shown promising results for learning the text and label representation in the same latent space. We further the research by proposing a novel architecture to derive superior contextual text representations using state-of-the-art language model BERT for multi-label defect triage.

Most of these published research benchmarks are on open-source defect report datasets - Eclipse and Mozilla \citep{6624028}. However, these datasets are focused on technical errors generated during system failures and do not mimic our use case. Our product defects are comprehensive user testing reviews consisting of natural language, technical and domain-specific text. In the real world, gathering labeled data is hard and expensive. Hence, we propose a methodology to generate a robust proprietary multi-label training dataset using weak supervision and adversarial learning.

\section{Data}
Our primary dataset is a proprietary in-house dataset consisting of actual defect reviews generated by beta testers for one of our major software releases. We rely on defect title and description fields to create the text corpus and text labels to identify the teams uniquely. Each defect could have multiple associated teams and vice versa. For our research, we have 3485 samples as a train set and 85 samples as the test set with 15 unique team labels for our multi-label dataset. Refer Table~\ref{tab:defect text}. We have 4-5 human-expert annotated defects corresponding to each team label in our low-resource setting. Our data preparation pipeline follows the below steps,

\begin{table*}
\centering
\begin{tabular}{lll}
\hline
\hspace{4cm}\textbf{Defect Text Corpus (Anonymized Excerpts)} \\
\hline
\verb|...For a store only query like XXX i am seeing available for| \\ \verb|scheduled pickup as the stack title on FE when i don't have a| \\
\verb|slot booked.This stack title should just reflect the XXX query| \\
\verb|like ios and web..Incorrect XXX mapping (number mapped to XXX..| \\
\hline
\verb|...I cant add XXX to my cart from order details from my previous| \\ \verb|canceled order. There is no actionable CTA.There is an add to cart|\\ \verb|CTA for the XXX. See attached video. Using ios XXX...| \\
\hline
\end{tabular}
\caption{\label{citation-guide}
Samples of Defect Text Corpus.
}
\label{tab:defect text}
\end{table*}

\subsection{Generate Labeled Data Using Weak Supervision}
Despite the success of fine-tuning pre-trained language models, one bottleneck is the requirement of labeled data. These labeled training data were expensive and time-consuming to create. It required human annotators with domain expertise to read through each defect review and assign team labels accordingly. Every change in labeling guidelines, team orientation, or use case changes necessitated re-labeling. Hence, we used Snorkel label model \citep{DBLP:journals/corr/abs-1711-10160} to generate weak labels for our training data. We apply 25 labeling functions (LFs) to unlabelled training data using a snorkel pipeline. Refer Table~\ref{tab:snorkel rules}.

\begin{table}[h!]
    \centering\small
    \begin{tabular}{|c|c|}
        \hline
        Rule & corpus->labels (Anonymized)\\
        \hline
        Keyword & 'android' or 'ios' -> [Team-LabelA] \\
        Pattern & '*search*' -> [Team-LabelB, Team-LabelC] \\
        \hline
    \end{tabular}
    \caption{Example LFs For Snorkel pipeline}
    \label{tab:snorkel rules}
\end{table}

\subsection{Generate Synthetic Data Using Adversarial Learning}
Machine learning algorithms are often vulnerable to adversarial examples that have imperceptible alterations from the original counterparts but can fool the state-of-the-art models \citep{DBLP:journals/corr/abs-1907-11932,DBLP:journals/corr/abs-2112-11668}.To increase the robustness, model training can be done using adversarial examples ~\citep{goodfellow2014generative, DBLP:journals/corr/abs-2110-09468}. We use Textattack framework \citep{DBLP:journals/corr/abs-2005-05909} on 30\% of our data, chosen at random to generate synthetic data for training our models and append these synthetic examples to our train set. We use embedding recipe of the framework that augments text by replacing words with neighbors in the counter-fitted embedding space, with a constraint to ensure their cosine similarity is at least 0.8. For every sampled defect, we produce 2 augmented defect texts by altering 10\% of original text words, while preserving the team labels . Refer Table~\ref{tab:adverserial text}


\begin{table*}
\centering
\begin{tabular}{lll}
\hline
\textbf{Defect Text (Anonymized)} & \textbf{Adversarial Defect Text (Anonymized)}\\
\hline
Price \underline{showing} inconsistently & Price \underline{displaying} inconsistently \\
Final \underline{cost} by weight not showing on search tiles & Final \underline{prices} by weight not showing on search tiles \\
Spacing on \underline{Nutrition} Label is too large & Spacing on \underline{Nourishment} Label is too large \\
\hline
\end{tabular}
\caption{\label{citation-guide}
Sample cases of Defect text vs Adversarial Defect Text.
}
\label{tab:adverserial text}
\end{table*}

\subsection{Fix Data Imbalances}
We found that the final training data created using the above techniques were imbalanced.
This issue was because the product defects were likely skewed towards a specific defect associated with a more significant and frequently tested domain vs. a rarely occurring one. We also noticed that defect reviews for features related to new team labels are getting introduced into the environment on an ongoing basis. To resolve the skewness, we used Multilabel Synthetic Minority Over-sampling Technique (MLSMOTE) \citep{CHARTE2015385} w.r.t the team labels with minimal data representation.

\begin{figure*}[h!]
    \centering
    \includegraphics[scale=0.35]{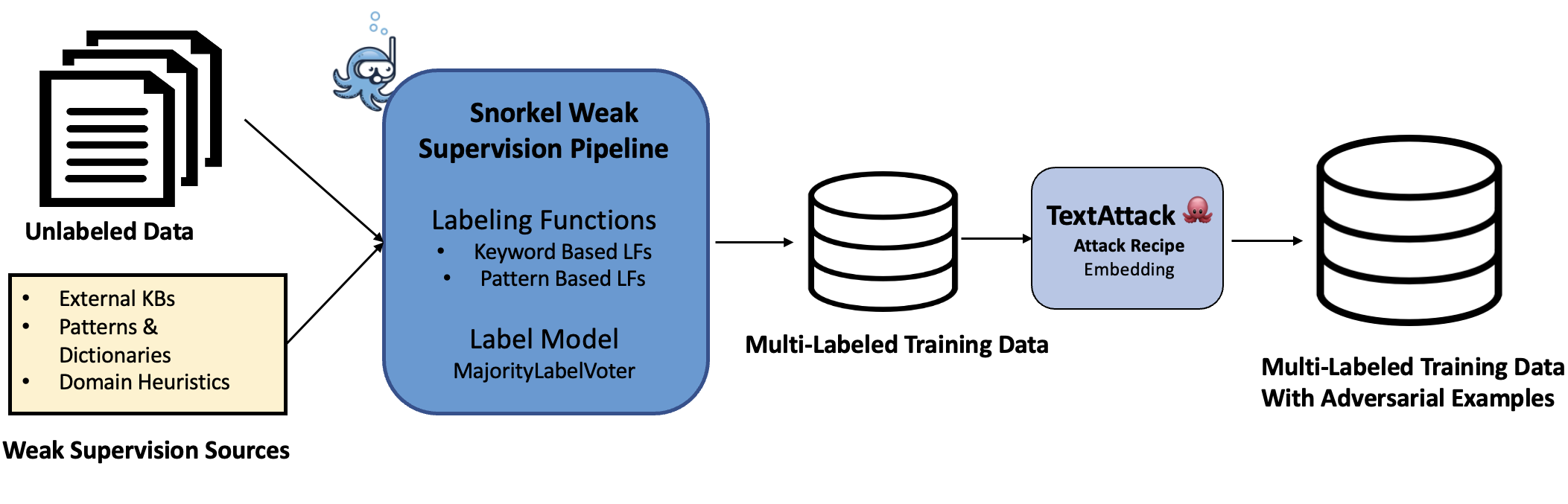}
    \caption{DEFTri Data Generation Methodology}
    \label{fig:otherdata2}
\end{figure*}

\section{Model}
The multi-team-labels defect classification task in this research can be summarized with S as the tuple set. $d_i$ and $t_i$ represents the $i^{th}$ defect denoted as D and its corresponding team-labels denoted as T. N, n and m are the total number of defects, the length of the $i^{th}$ defect text and the number of teams-labels of the $i^{th}$ document, respectively.
\begin{math}
\\
\\
 S = \{(d_i, t_i)\}_{i=1}^N, D = \{d_i|d_i = \{d_1, d_2,· · · , d_n\}\},
 \\
T=\{t_i|t_i =\{t_1,t_2,···,t_m\}\}
\\
\\
\end{math}
Our framework, DEFTri aims at assigning team-labels to its corresponding defects based on the conditional probability P($t_i$|$d_i$).

\subsection{Pre-Trained Model}
For our fine-tuning, we use BERT pre-trained transformer embedding from Hugging Face’s Transformers library ~\citep{wolf2020huggingfaces}.BERT base uncased embeddings are case insensitive and are pre-trained on the English language self-supervised using two objectives - masked language modeling (MLM) and Next Sentence Prediction (NSP). These embeddings were introduced in the original BERT ~\citep{DBLP:journals/corr/abs-1810-04805} paper and serve as baseline embeddings for our models.

\subsection{Approach}
For our DEFTri framework, we propose 2 novel implementations to derive superior contextual representations from product defect text, that help in improved multi-label defect classification task. We denote the defect corpus(title and description) tokens as $D_i$ and their corresponding token embeddings as $E_{Di}$, where K is the total number of words in the input defect and $D_K$ represents the last token. Similarly, let $L_j$ be the team label text of the $j^{th}$ team of the overall 15 teams, corresponding to the defect corpus. Finally, we derive the positional embedding using BERT and apply classification layer with activation to the last layer of the hidden state at the [CLS] token.

\subsubsection{Label Fused Model with [SEP]}
\begin{figure}[h!]
    \centering
    \includegraphics[scale=0.25]{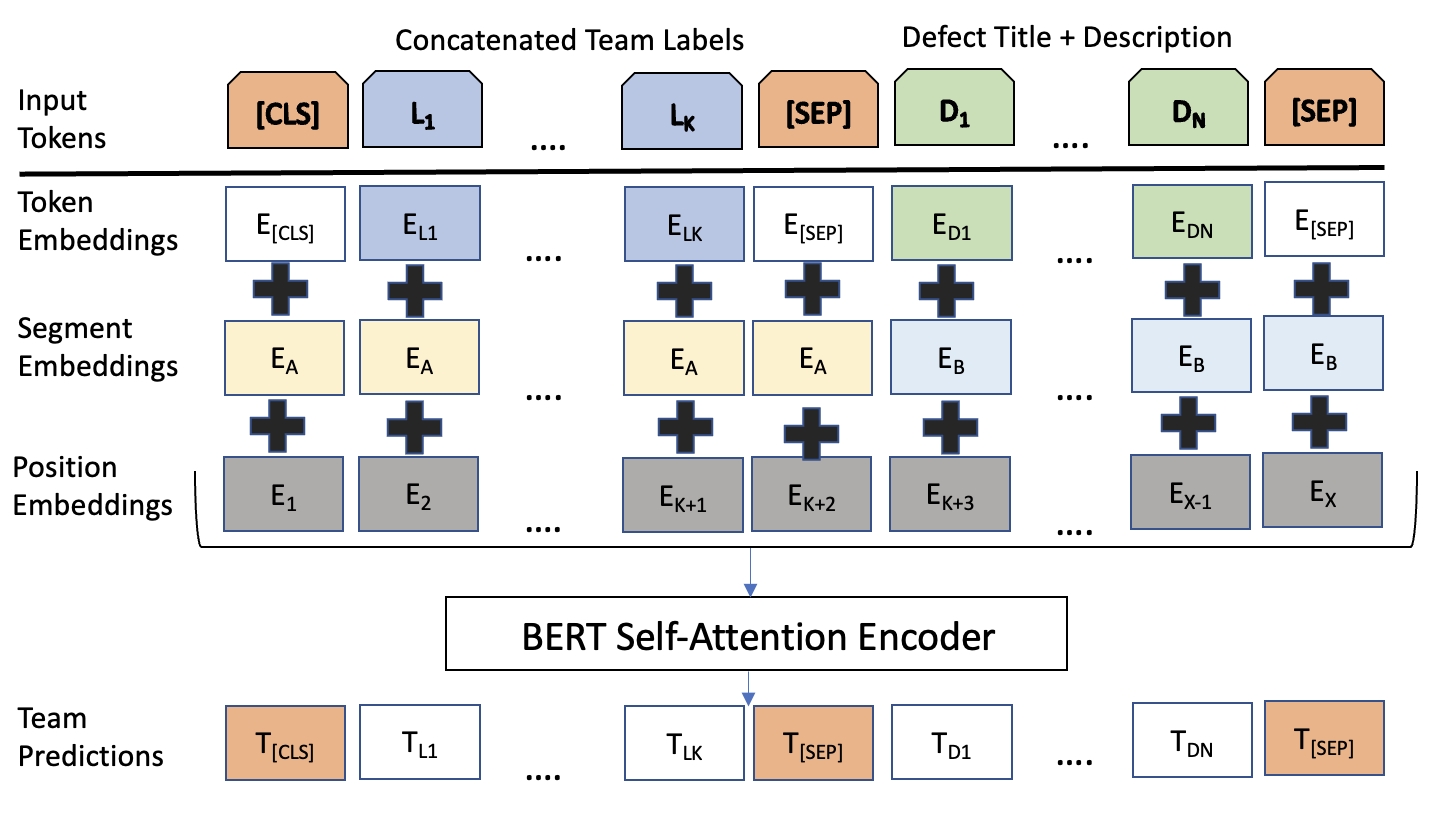}
    \caption{DEFTri LabelFuse Model with [SEP]}
    \label{fig:modelarchitecture1}
\end{figure}

We utilize the sentence pair configuration of BERT for text input. We concatenate the team labels text as Sentence A and concatenate the Defect title and description text as Sentence B, both separated by a [SEP] token. Refer Figure~\ref{fig:modelarchitecture1}

\subsubsection{Label Fused Model without [SEP]}
\begin{figure}[h!]
    \centering
    \includegraphics[scale=0.25]{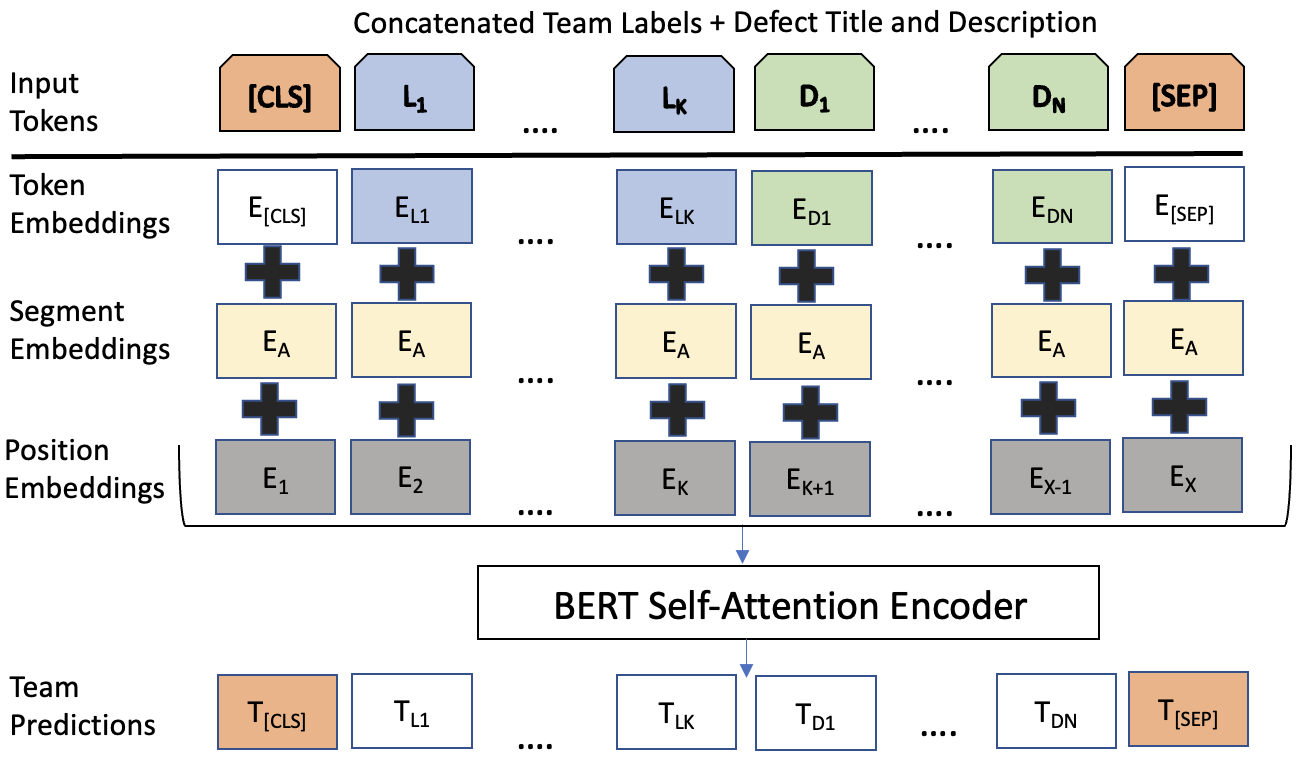}
    \caption{DEFTri LabelFuse Model w/o [SEP]}
    \label{fig:modelarchitecture2}
\end{figure}

For our second implementation, we concatenate the team labels text along with Defect title and description text as a single Sentence A, without any [SEP] token as input. Refer Figure~\ref{fig:modelarchitecture2}

\begin{table*}[h!]
\centering
\begin{tabular}{lll}
\hline
\textbf{Model} & \textbf{Macro-F1} & \textbf{Accuracy}\\
\hline
BERT+Linear &  0.8123 & 0.8134 \\
BERT+BiLSTM &  0.8206 & 0.8216 \\
\hline 
BERT+LabelFuse w/o [SEP]+Linear & 0.8144 & 0.8153 \\
BERT+LabelFuse w/o [SEP]+BiLSTM & \textbf{0.8236} & \textbf{0.8245} \\
\hline
BERT+LabelFuse w [SEP]+Linear & 0.8137 & 0.8150 \\
BERT+LabelFuse w [SEP]+BiLSTM & 0.8229 & 0.8241 \\
\hline
\end{tabular}
\caption{
DEFTri Experiments Results For Contextual Multi-TeamLabel Classification on Real Product Defects
}
\label{tab:experimenttable}
\end{table*}

\subsection{Classification Head}
We experimented with two different dense layers for the classification head - Linear and BiLSTM. Refer Table~\ref{tab:classificationhead} 

\begin{table}[h!]
    \centering\small
    \begin{tabular}{|c|c|c|c|c|c|c|}
        \hline
        \multicolumn{2}{|c|}{Classification Heads} & \multicolumn{2}{|c|}{Dense Layer} & \multicolumn{2}{c|}{Linear Layer} \\
        \hline
        Type & Activation & In & Out & In & Out \\
        \hline
        Linear & Tanh & 768 & 768 & 768 & 15 \\
        BiLSTM & ReLU & 768 & 256 & 512 & 15\\
        \hline
    \end{tabular}
    \caption{DEFTri Classification Head Configurations}
    \label{tab:classificationhead}
\end{table}

\subsection{Loss Function and Optimizer}
For model training we use PyTorch implementation of BCEWithLogitsLoss as our loss function and AdamOptimizer as our optimizer. BCEWithLogitsLoss combines a Sigmoid layer and the Binary Cross Entropy Loss in one single class. In case of multi-label classification the loss can be described as,

\begin{math}
\\
l_t(x,y) = L_t = \{l_{1,t},...,l_{N,t}\}^T,
\\
l_{n,t} = -w_{n,t}[p_ty_{n,t} \cdot log\sigma(x_{n,t}) + 
(1+y_{n,t}) \cdot log\sigma(x_{n,t})]
\end{math}

where t=15 and represents the number of team-labels , n is number of sample in the batch and $p_t$ is the weight of the positive answer for team-label t. 

\subsection{Hyper-Parameters}
We use a set of hyper-parameters for our experiments. We used manual search for hyper-parameter search and the best model was chosen based on the best top-1 accuracy yielded in the validation data. Refer Table~\ref{tab:hyperparams}

\renewcommand{\arraystretch}{1}
\begin{table}[h!]
    \centering\small
    \begin{tabular}{|c|c|}
        \hline
        HParams & Values \\
        \hline
        Dropout & 0.1 \\
        Max Sequence Length & 512 \\
        Batch-Size & 16 \\
        Learning Rate & 1e-5 \\
        Weight Decay & 0.01 \\
        Adam epsilon & 1e-6 \\
        Epochs & 10 \\
        \hline
    \end{tabular}
    \caption{DEFTri Hyper-Parameters}
    \label{tab:hyperparams}
\end{table}

\section{Experiments}
As baseline and our proposed architecture, we use the pre-trained bert-base-uncased model ~\citep{wolf2020huggingfaces, DBLP:journals/corr/VaswaniSPUJGKP17}. We perform a total of 6 experiments for our models under 3 different settings (1) baseline fine-tuned BERT model with no fused labels (2) fine-tuned BERT with fused labels without [SEP] token and (3) fine-tuned BERT with fused labels with [SEP] token, using 2 classification heads combinations e.g Linear and BiLSTM. Refer Table~\ref{tab:experimenttable} and Appendix~\ref{sec:appendixexperimentsetting}

For data preprocessing step, the corpus is converted to lowercase and tokenzied with one-hot-encoded labels.Our deep learning model is then trained to predict multiple team-labels for each test sample. At inference time, the model takes in an input of text corpus of defect and predicts a vector of probabilities  for each of the 15 team-labels. We used a confidence threshold of 0.55 for our probability vector to obtain a binary vector for comparison with ground-truth.

Measuring accuracy on exact binary vector matching for multi-label classification is too penalizing because of the low tolerance for partial errors. Therefore, we divide our predictions by classes. For each of the team-labels in our dataset, we calculate the number of false positives (FP), false negatives (FN), true positives (TP), true negatives (TN). Finally, to obtain our Accuracy, we sum up the values across each team-labels as below,
\begin{math}
\\
\\
Accuracy=\frac{\sum TP_t+\sum TN_t}{\sum FP_t+\sum FN_t+\sum TP_t+\sum TN_t}
\\
\\
\end{math}
where T=15 and represents the number of team-labels in our dataset and $TP_t$, $TN_t$, $FP_t$ , $FN_t$ represents values of TP, TN, FP, FN for $t^{th}$ team-label. Similarly, we used macro-F1 (F1) scores based on averaged value of precision and recall calculated over all team-labels as below,
\begin{math}
\\
\\
Precision_t = \frac{TP_t}{FP_t + TP_t}
\\
\\
Recall_t = \frac{TP_t}{FN_t + TP_t}
\\
\\
F1 = 2 \times \frac{\frac{1}{T} \sum Precision_t \times \frac{1}{T} \sum Recall_t}{\frac{1}{T}\sum Precision_t + \frac{1}{T}\sum Recall_t }
\\
\end{math}
\section{Analysis}
Based on our experiments, we observed that label-fused contextual learning-based fine-tuned BERT models significantly outperformed the base model using only the context of the defect text. The performance boost over the base BERT pre-trained fine-tuned model is because of the context in the label embeddings used in addition to the defect text in the label-fused models, which optimizes on the alignment of features, which makes it possible to classify better. Our team labels were short meaningful English words vs abbreviations which made fused embeddings better for classification when paired as a sentence with the defect texts as inputs. We observed that label-fused model without [SEP] token performed better that with [SEP] token which could have been because of the unnatural formation of Sentence A, where a bunch of team labels are concatenated together.  

Also, with the addition of synthetically generated data using adversarial examples for model training, we achieved an average accuracy improvement of 2.69\% across our models vs. using the original data only. However, during our experiments we observed that the performance was sensitive towards the choice of text corpus sequence length and perturbation percentage for data augmentation made, during model training. A higher percentage of perturbations combined with a lower sequence length of text corpus negatively impacted performance.

\section{Future Work}
Fine-tuning language models with weak supervision definitely solves the challenge of low labeled data availability. However, the models performance definitely suffers from error-propagation of pseudo-labels generated during the process. Recent research in contrastive self-regularized self-training approach \citep{DBLP:journals/corr/abs-2010-07835} and GAN-BERT in adversarial setting \citep{croce-etal-2020-gan}  have shown promising results for fine-tuning BERT-based language models with weak supervision. Also, Contrastive learning and Adversarial Learning approaches applied to various NLP tasks have demonstrated improvement over fine-tuning on BERT-based models \citep{DBLP:journals/corr/abs-2112-01054,DBLP:journals/corr/abs-2107-10137}. To further our research, we would improve upon these approaches.

\section{Conclusion}
In this work, we proposed a novel framework, DEFTri for automated defect triage using contextual representations of human-generated defect reviews at Walmart. We discussed our methodology of generating a new proprietary labeled dataset by using weak supervision and adversarial learning, in a few shot setting. We presented two label-fused model approaches for fine-tuning pre-trained BERT. As hypothesized, the experimental results show that our approach improves the multi-label text classification task for defect triage. We also proposed our future work of implementing contrastive learning for fine-tuning using weak supervision. 

\section*{Acknowledgments}
The author would like to thank colleagues in the Omni Customer Experience org. at Walmart Global Tech for all their support and encouragement. 

\bibliography{anthology,custom}
\bibliographystyle{acl_natbib}

\appendix

\section{Appendix}
\label{sec:appendix}

\subsection{Experiment Setting}
\label{sec:appendixexperimentsetting}
 We ran all our experiments on a Google Cloud Platform using a n1-standard-16 machine with NVIDIA Tesla V100 GPUs.

\end{document}